\documentclass{aa}
\usepackage{graphics}
\newcommand{\phminus}{\phantom{-}}

\begin{document}

   \title{Radio continuum spectra of galaxies in the Virgo cluster
          region\thanks{Based on observations with the 100-m
          telescope of the MPIfR (Max-Planck-Institut f\"{u}r
          Radioastronomie) at Effelsberg}   }

   \author{B.~Vollmer\inst{1,2}, M.~Thierbach\inst{2}, \& R.~Wielebinski\inst{2}}

   \offprints{B.~Vollmer, e-mail: bvollmer@astro.u-strasbg.fr}

   \institute{CDS, Observatoire astronomique de Strasbourg, UMR 7550, 11, rue de l'universit\'e,
     67000 Strasbourg, France \and
     Max-Planck-Institut f\"ur Radioastronomie, Auf dem H\"ugel 69, 53121 Bonn Germany
   }

   \date{Received / Accepted}

   \authorrunning{Vollmer et al.}
   \titlerunning{Virgo radio continuum spectra}

\abstract{
New radio continuum observations of galaxies in the Virgo cluster region at 4.85, 8.6, and
10.55~GHz are presented. These observations are combined with existing measurements
at 1.4 and 0.325~GHz. The sample includes 81 galaxies were spectra with more than two
frequencies could be derived. Galaxies that show a radio--FIR excess exhibit central activity
(HII, LINER, AGN). The four Virgo galaxies with the highest absolute radio excess are found
within 2$^{\rm o}$ of the center of the cluster. Galaxies showing flat radio spectra also
host active centers. There is no clear trend between the spectral index and the galaxy's distance
to the cluster center.
}

\maketitle

\section{Introduction \label{sec:intro}}

A galaxy cluster is an ideal laboratory for studying the influence of the galaxies'
environment on its appearance and/or evolution.
It is a well established fact that spiral galaxies in clusters have less
atomic gas than isolated spirals of the same morphological type and same
optical diameter, i.e. they are H{\sc i} deficient (Chamaraux et al. 1980,
Bothun et al. 1982, Giovanelli \& Haynes 1985, Gavazzi 1987, 1989).
There are mainly two kinds of mechanisms which are able to cause the removal
of the atomic gas: (i) tidal interactions or (ii) the interaction of the
interstellar medium (ISM) with the hot intracluster medium (ICM). The mapping of
the gas content of spiral galaxies in the Virgo cluster (Cayatte et al. 1990, 1994)
showed that the H{\sc i} disk sizes of cluster
spirals are considerably reduced. In addition, galaxies with a symmetric
optical disk that have an asymmetric H{\sc i} gas distribution are quite
frequent in the cluster core. These observational results indicate that
the gas removal due to the rapid motion of the galaxy within the ICM
(ram pressure stripping; Gunn \& Gott 1972) is responsible for the H{\sc i}
deficiency and the distorted gas disks of the cluster spirals.

During the phase of active ram pressure stripping the interstellar matter (ISM), which is located
in the outer disk is pushed to smaller radii where it is compressed (Vollmer et al. 2001).
This effect is more important for edge-on stripping than for face-on stripping.
Since the magnetic field is frozen into the ISM, the magnetic field $B$ becomes compressed too
and its strength is enhanced (Otmianowka-Mazur \& Vollmer 2003).
The radio continuum emission due to relativistic particles gyrating
around the magnetic fields is proportional to $B^{2-4}$ depending on the fact if there is
equipartition between the relativistic electrons and the magnetic field.
One might thus expect an enhancement of the radio continuum emission of spiral galaxies that undergo
active ram pressure stripping. This enhancement should be detectable as a radio excess in the well
studied radio--FIR correlation (see, e.g., de Jong et al. 1985, Wunderlich et al. 1987, Niklas 1997).

Furthermore, one might speculate that ram pressure stripping triggers nuclear activity.
Since radio spectra of centrally active galaxies (HII, LINER, AGN) are generally flatter than
the synchrotron spectra of normal galaxies (mean spectral index $\alpha \sim -0.7$,
Israel \& van der Hulst 1983), one might expect flatter radio spectra close to the
cluster center.

Gavazzi \& Boselli (1999a) studied the radio luminosity function of Virgo cluster galaxies for
early and late type galaxies separately. They found that late type galaxies develop radio
sources with a probability proportional to their optical luminosity, independently of
their detailed Hubble type. In a second paper Gavazzi \& Boselli (1999b) compared the radio
luminosity functions of galaxies in different clusters to those of isolated galaxies.
They concluded that the radio luminosity function of Virgo cluster galaxies is consistent
with that of isolated galaxies, whereas the Coma cluster galaxies show an excess of
radio emissivity.

Niklas et al. (1995) investigated the behaviour of the radio--FIR correlation and the radio
spectra in the Virgo cluster. They claim that galaxies that show an excess of radio emission
are located close to the cluster center, whereas the spectra do not seem to be affected by
the location of the galaxies in the cluster.

In this article we extend the galaxy sample of Niklas et al. (1995) in adding observations at 8.6~GHz and in
enlarging the sample to 81 galaxies, which were observed at 10.55, 8.6, 4.85~GHz with the Effelsberg
100-m telescope. Our data is complemented with 1.4~GHz flux densities from the NVSS where advantage is
taken of the NVSS images to check for source confusion and to estimate the correct source flux.

This article has the following strucutre:
The sample and the observations are presented in Sec.~\ref{sec:sample} and Sec.~\ref{sec:obs}.
We show the results in Sec.~\ref{sec:results} and investigate the radio--FIR correlation
in Sec.~\ref{sec:radiofir}. The distribution of the radio spectra are studied in Sec.~\ref{sec:specindex}
followed by the discussion and our conclusions (Sec.~\ref{sec:discon}).

\section{The sample \label{sec:sample}}

\begin{figure}
        \resizebox{\hsize}{!}{\includegraphics{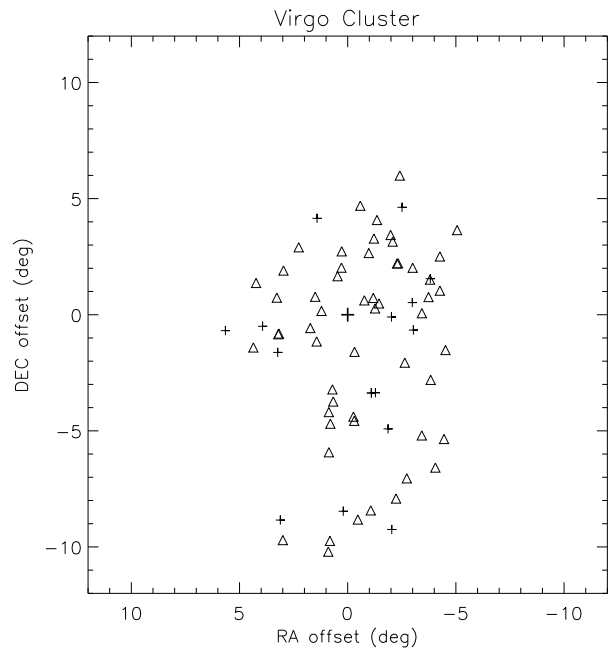}}
        \caption{Distribution of the observed galaxies. Triangles: Virgo galaxies.
   Crosses: background galaxies. Thick cross: M87.
        } \label{fig:virgo_map}
\end{figure}

Gavazzi \& Boselli (1999a) derived the radio luminosity function of the Virgo cluster
by cross-correlating optical data of the Virgo Cluster Catalogue (VCC) (Binggeli et al. 1985)
with NVSS radio survey carried out at the VLA at 1.4~GHz (Condon et al. 1998).
The 180 positive radio-optical matches that they found build the bases of our survey.
We restrict ourselves to radio sources with 1.4~GHz flux densities greater than 10~mJy, which
corresponds to a flux density at 10.5~GHz of 2~mJy, assuming a spectral index $\alpha=-0.8$, where
\begin{equation}
S \propto \nu^{\alpha}
\label{eq:specindex}
\end{equation}
 ($S$ is the radio flux density and $\nu$ the frequency).
We thus observed 81 galaxies of various Hubble types. Their distribution on the sky
is shown in Fig.~\ref{fig:virgo_map}. The number of galaxies of different
types are listed in Tab.~\ref{tab:types}. As expected, the sample is clearly dominated by
late type galaxies (Sc/Scd) with a high star formation rate. There are also 10 dwarf galaxies in
our sample. We took the $\lambda$20, 50~cm flux densities, and the
IRAS FIR fluxes from the Goldmine database\footnote{This research has made
use of the GOLDMine Database, which is operated by the University of Milano-Bicocca
(see Gavazzi et al. 2003, A\&A, 400, 451)} and the position angle of the galactic disks from LEDA
\footnote{HyperLeda is an extragalactic database located at http://leda.univ-lyon1.fr/}.
The FIR flux is calculated according to Helou et al. (1988) in the following way:
\begin{equation}
FIR=2.58 S_{60_{\mu}} + S_{100_{\mu}}\ ,
\label{eq:fir}
\end{equation}
where $S_{60_{\mu}}$ is the flux at 60~$\mu$m and $S_{100_{\mu}}$ the flux at 100~$\mu$m.

\begin{figure}
        \resizebox{\hsize}{!}{\includegraphics{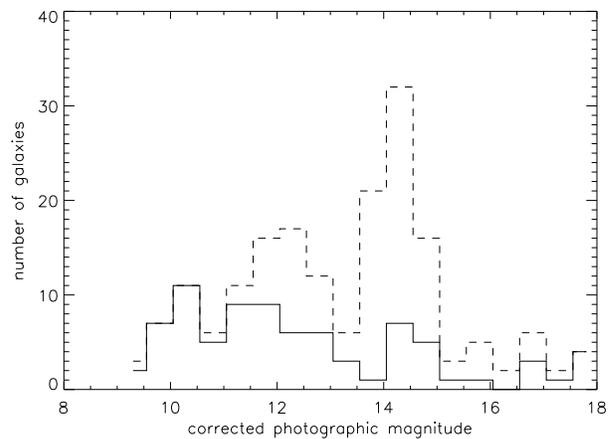}}
        \caption{Number of galaxies as a function of the corrected B band magnitude in bins of
   0.5~mag. Solid: our sample. Dashed: complete sample of Virgo galaxies brighter than
   m$_{\rm B}$=18 (Gavazzi \& Boselli 1999a).
        } \label{fig:complete}
\end{figure}

Fig.~\ref{fig:complete} shows the number of galaxies as a function of the corrected B band
magnitude in bins of 0.5~mag. For comparison, the complete sample
of Virgo galaxies brighter than m$_{\rm B}$=18 (Gavazzi \& Boselli 1999a) is also shown.
Most of the missing galaxies in our sample compared to that of Gavazzi \& Boselli (1999a)
have photographic magnitudes around m$_{\rm B}$=12 and m$_{\rm B}$=14.

\begin{table}
      \caption{The galaxy sample: number of galaxies per Hubble type.}
         \label{tab:types}
      \[
         \begin{array}{|c|c|c|c|c|c|c|c|c|}
         \hline
  {\rm E/S0} & {\rm S} & {\rm Sa/Sab} & {\rm Sb/Sbc} & {\rm Sc/Scd} & {\rm Sm/Im} & {\rm Pec} & {\rm dE} & {\rm BCD} \\
  \hline
  12 & 2 & 10 & 10 & 33 & 2 & 2 & 8 & 2 \\
  \hline
   \end{array}
      \]
\end{table}

\section{Observations \label{sec:obs}}

Our sample of 81 galaxies was observed November 2000, January 2001, August/September/November 2002, and
April/June 2003 with the 100-m Effelsberg telescope at 4.85, 8.6, and 10.55~GHz (6.0, 3.6, and 2.8cm).
The frequency, HPBW, system temperature, bandwidth, and number of beams are listed in Tab.~\ref{tab:obsc}.
\begin{table}
      \caption{Observations characteristics}
         \label{tab:obsc}
      \[
         \begin{array}{|c|c|c|c|c|}
         \hline
  {\rm Frequency} & {\rm HPBW} & T_{\rm sys}  & {\rm bandwidth} & {\rm beams} \\
  {\rm (GHz)} & {\rm ('')} & {\rm (K)}  & {\rm (GHz)} & \\

 \hline
   4.85 & 147 & 9 &  0.5 & 2 \\
 \hline
   8.6 & 85 & 4 & 1.2 & 1 \\
 \hline
   10.55 & 69 & 50 & 0.3 & 4 \\
  \hline
   \end{array}
      \]
\end{table}
The existence of several beams at 4.85 and 10.55~GHz
permits to subtract non-linear baselevels due to meteorologic conditions. The most sensitive
receiver is the one at 8.6~GHz due to its huge bandwidth and low system temperature. Since it is a
single beam receiver, perfect weather conditions are needed for observations at this frequency.

The galaxies were measured with cross-scans. The cross-scans were centered on the radio center of the
galaxies given by Gavazzi \& Boselli (1999a). The scanning directions were along the major and minor axis of the
galaxies. The length of the cross-scans were $16'$, $15'$, and $10'$, the duration of the scan
24, 30, and 30~sec at 4.85, 8.6, and 10.55~GHz respectively. The number of cross-scans and thus the
total integration time at a given frequency was calculated using the expected flux density extrapolated
from the NVSS flux at 1.4~GHz ($\alpha=-0.8$) and a S/N of 5.
The calibration scale for the fluxes is that of Ott et al. (1994).
For a correct calibration of the cross-scans and telescope pointing 3C286 was observed frequently
during the observations.

In order to derive the flux densities of the cross-scanned galaxies, Gaussians were fitted to the averaged
scans in each direction. These fits give the peak flux value $S^{\rm peak}$ and the half-power width
$HPW^{\rm obs}$ of the fitted Gaussian and the positional offset. The peak value of each scan direction
was then corrected using the following formula to obtain the total flux density $S_{\rm tot}$:
\begin{equation}
S_{\rm tot}=S_{1}^{\rm peak} \times S_{2}^{\rm peak} \times (\frac{HPW^{\rm obs}_{1}}{HPBW}) \times
(\frac{HPW^{\rm obs}_{2}}{HPBW})\ ,
\end{equation}
where $S_{i}^{\rm peak}$ and $HPW^{\rm obs}_{i}$ with $i=1, 2$ are the flux densities and half-power
widths along the major and minor axis of the galaxies.
For the determination of the source extent, the 10.55~GHz and 8.6~GHz data
are best suited, because they have the highest resolutions.
Unfortunately, meteorological conditions and frequent source confusion did us not allow to derive
reliable half-power widths from our data. Therefore, we used our peak fluxes and the
half power widths at 1.4~GHz given by Gavazzi \& Boselli (1999a).

Source confusion is a serious problem in the Virgo cluster region, especially at 4.85~GHz.
Our advantage is that we have the NVSS 1.4~GHz fluxes, where the correct source has been
identified on the NVSS images (Gavazzi \& Boselli 1999a). Thus, it was possible to identify
the confusing source, to subtract it, and to fit a Gaussian to the radio source belonging
to the correct VCC source. Since Niklas et al. (1995) did not have the NVSS flux densities
they could not decide which source was the correct one. Thus we had to re-observe several of
their measured galaxies.

\section{Results \label{sec:results}}

The observed flux densities at 4.85, 8.6, and 10.55~GHz are listed in Tab.~\ref{tab:parameters}.
We have also included the measurements of Niklas et al. (1995), which are marked in boldface.
The columns of Tab.~\ref{tab:parameters} are: (1) VCC name of the galaxy (2) right ascension in
B1950 coordinates (3) declination in B1950 coordinates (4) projected
distance from the Virgo cluster center
(M87) (5) Hubble type (6) Virgo cluster membership based on radial velocities (y/n) (7) flux
density at 10.55~GHz (8) flux density at 8.6~GHz (9) flux density at 4.85~GHz (10) flux density
at 1.4~GHz (11) flux density at 0.6~GHz.(12) FIR flux (see Eq.~{\ref{eq:fir}) (13) spectral index
$\alpha$ (see Eq.~\ref{eq:specindex}). When we did not detect a galaxy at 4.85, 8.6, and 10.55~GHz
the upper limit of the flux density is $\sim 1$~mJy at 10.55~GHz.

We made linear fits in the $\log\,\nu - \log\,S_{\nu}$ plane using the method of the least
absolute deviation. This method is less sensitive to outlying points in the spectrum and thus
more robust than a $\chi^{2}$ fit. The obtained spectra are shown in Fig.~\ref{fig:spectra}.
The spectral index $\alpha$ is plotted in the lower left corner of the boxes. We only fitted
a spectrum when there are more than two data points.

Since the cross-scan method is less accurate than imaging of the sources, the spectra
are noisier than those obtained by direct imaging. The main error is due to
offsets in the Gaussian fits and imprecise position angles of the galaxies.
In many cases the flux densities at 10.55~GHz are lower than expected by our fitted spectrum.
This might be due to the fact that this receiver is the least sensitive of the three receivers used.
Moreover, the background subtraction using a second beam has the effect that the total flux
might be underestimated.

\begin{figure*}
        \resizebox{14cm}{!}{\includegraphics{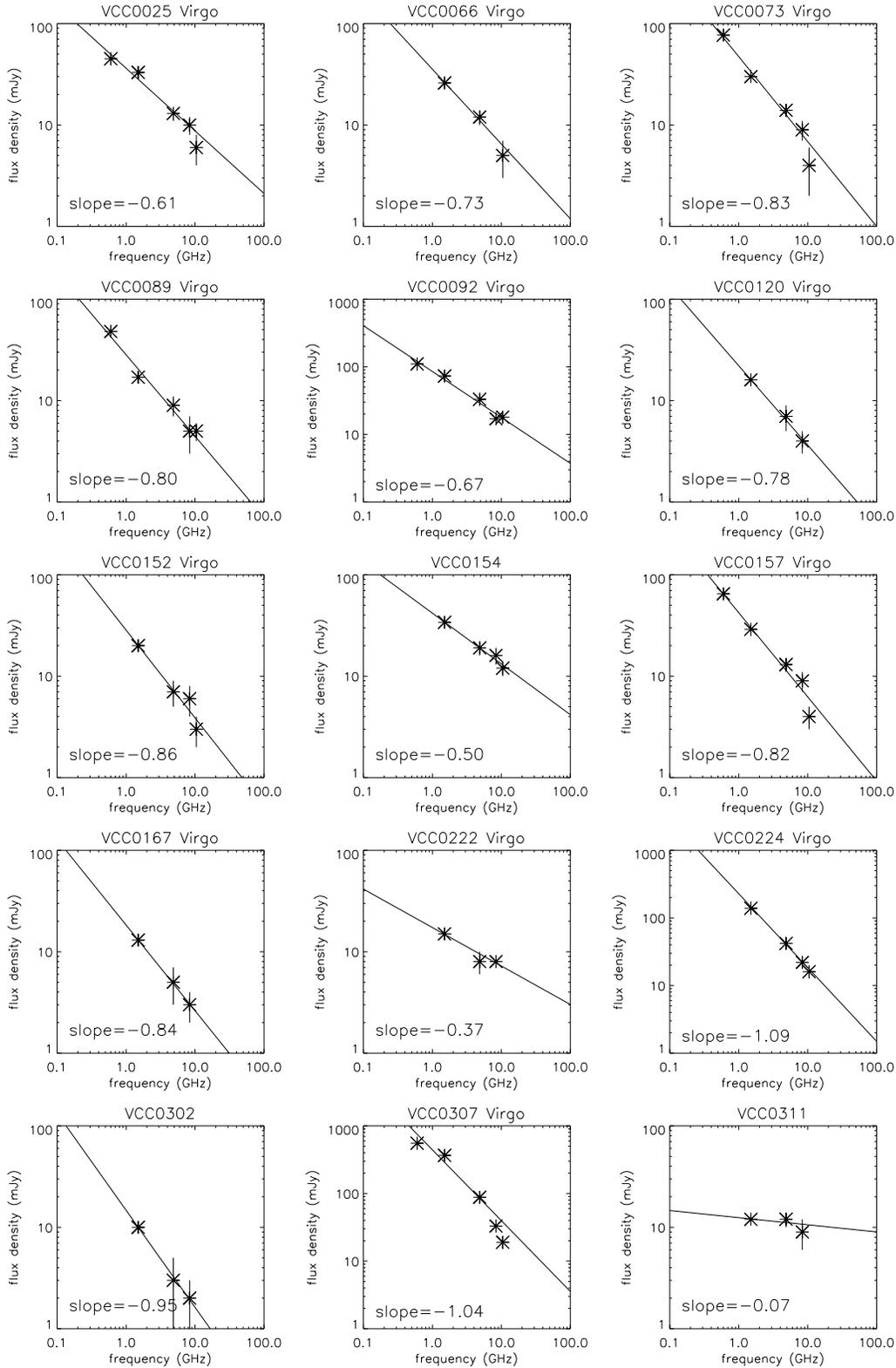}}
        \caption{Radio spectra of the observed VCC sources.
        } \label{fig:spectra}
\end{figure*}
\addtocounter{figure}{-1}
\begin{figure*}
        \resizebox{14cm}{!}{\includegraphics{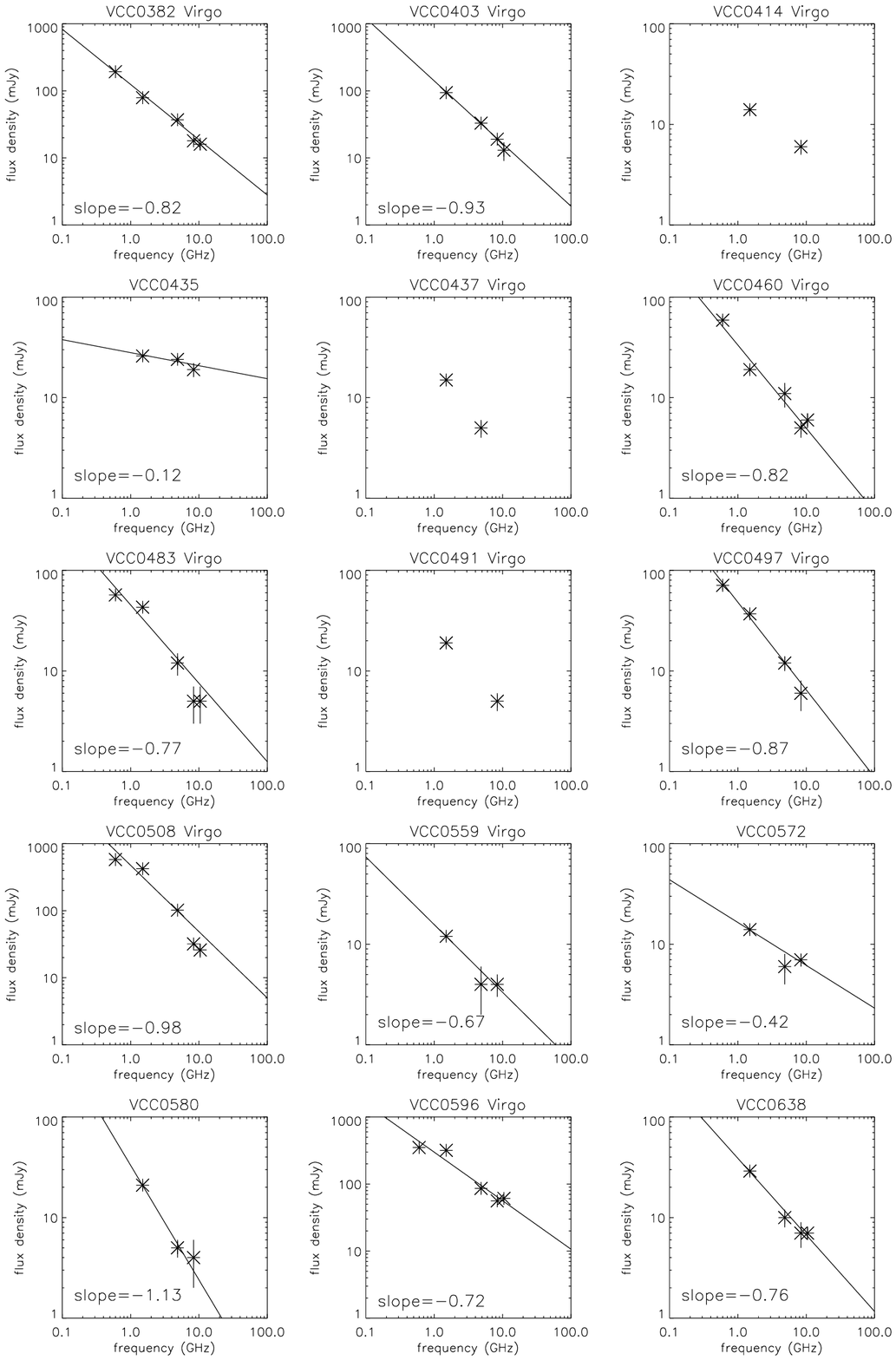}}
        \caption{Radio spectra of the observed VCC sources (continued).}
\end{figure*}
\addtocounter{figure}{-1}
\begin{figure*}
        \resizebox{14cm}{!}{\includegraphics{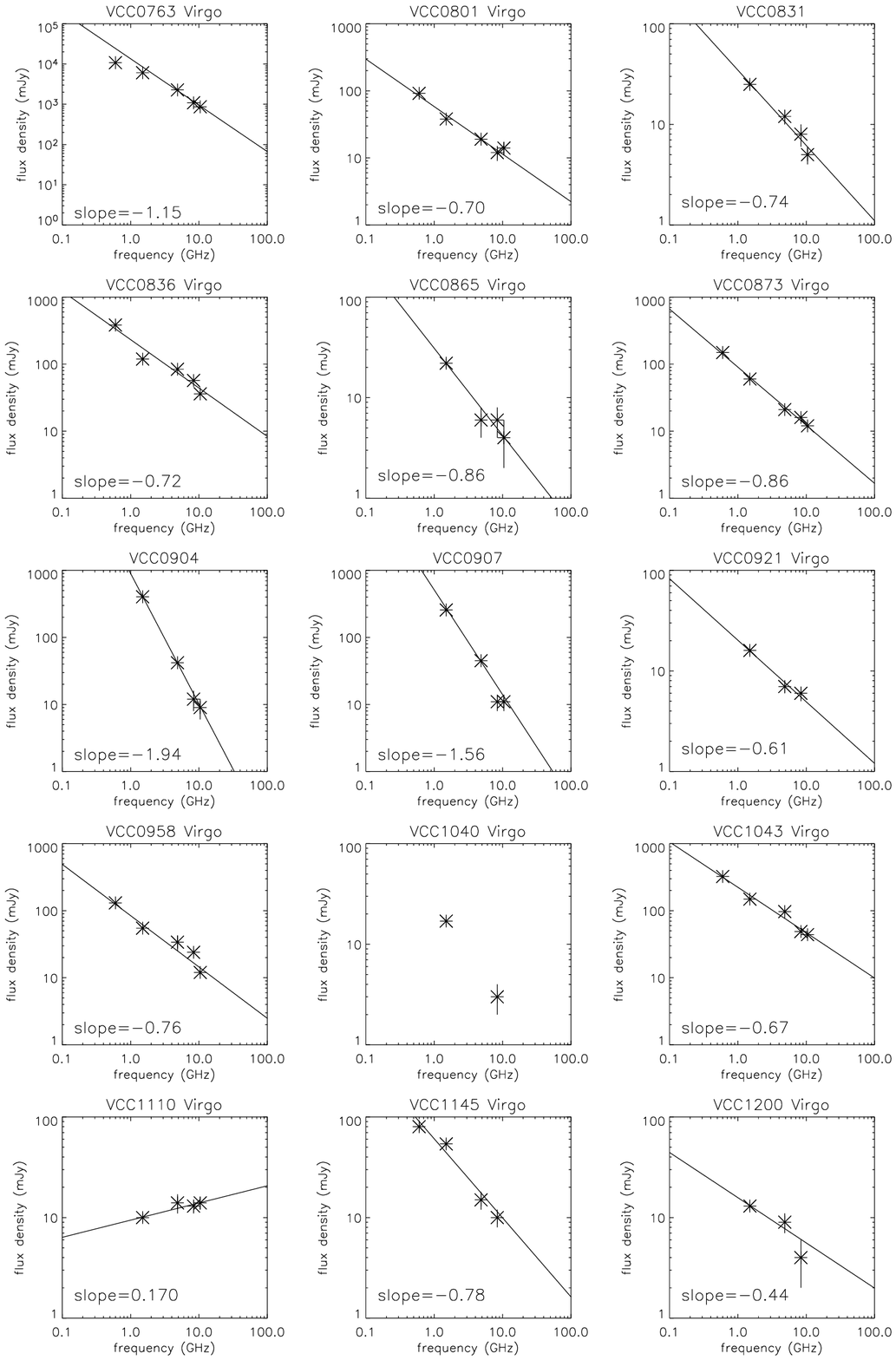}}
        \caption{Radio spectra of the observed VCC sources (continued).}
\end{figure*}
\addtocounter{figure}{-1}
\begin{figure*}
        \resizebox{14cm}{!}{\includegraphics{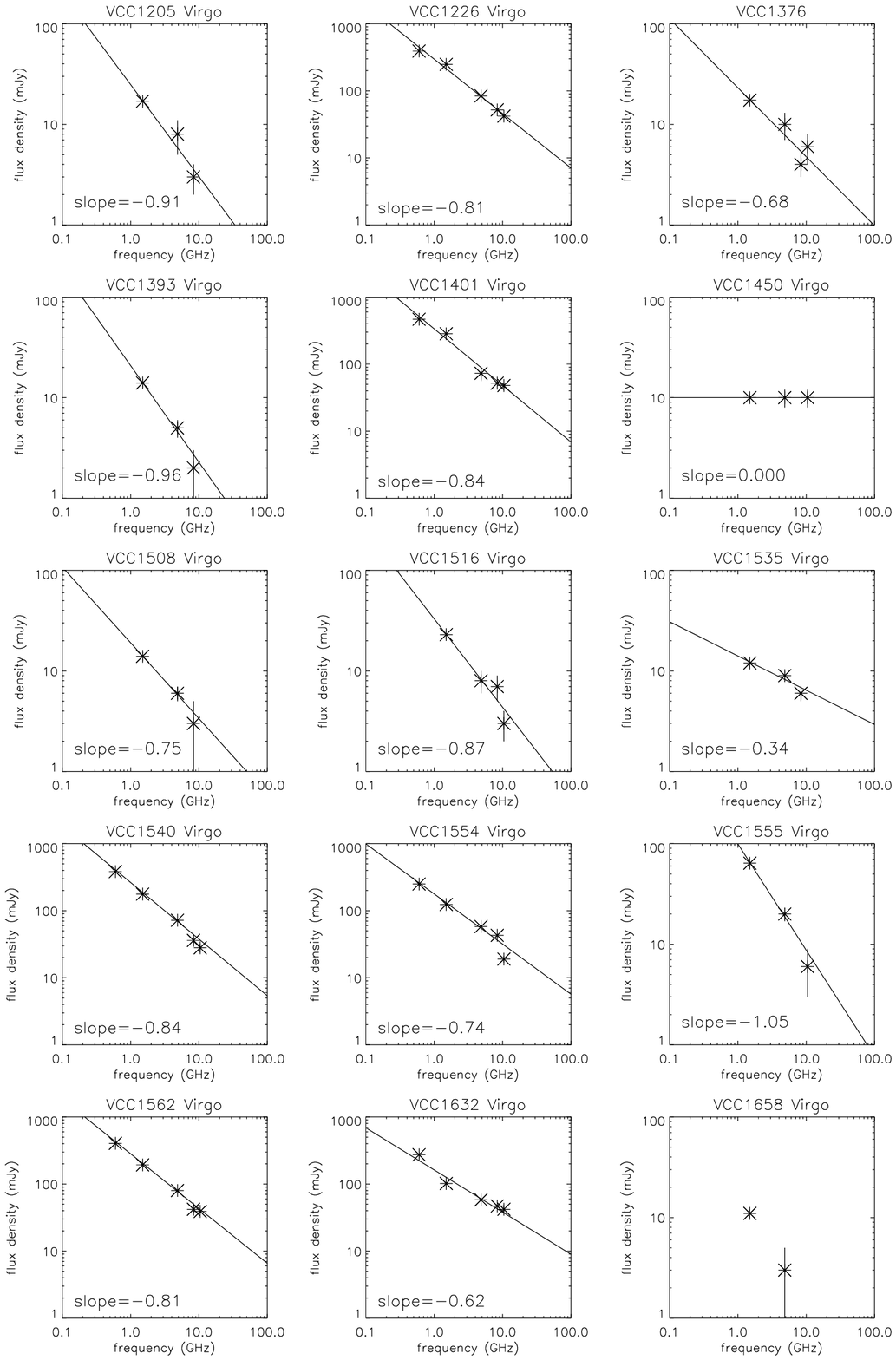}}
        \caption{Radio spectra of the observed VCC sources (continued). }
\end{figure*}
\addtocounter{figure}{-1}
\begin{figure*}
        \resizebox{14cm}{!}{\includegraphics{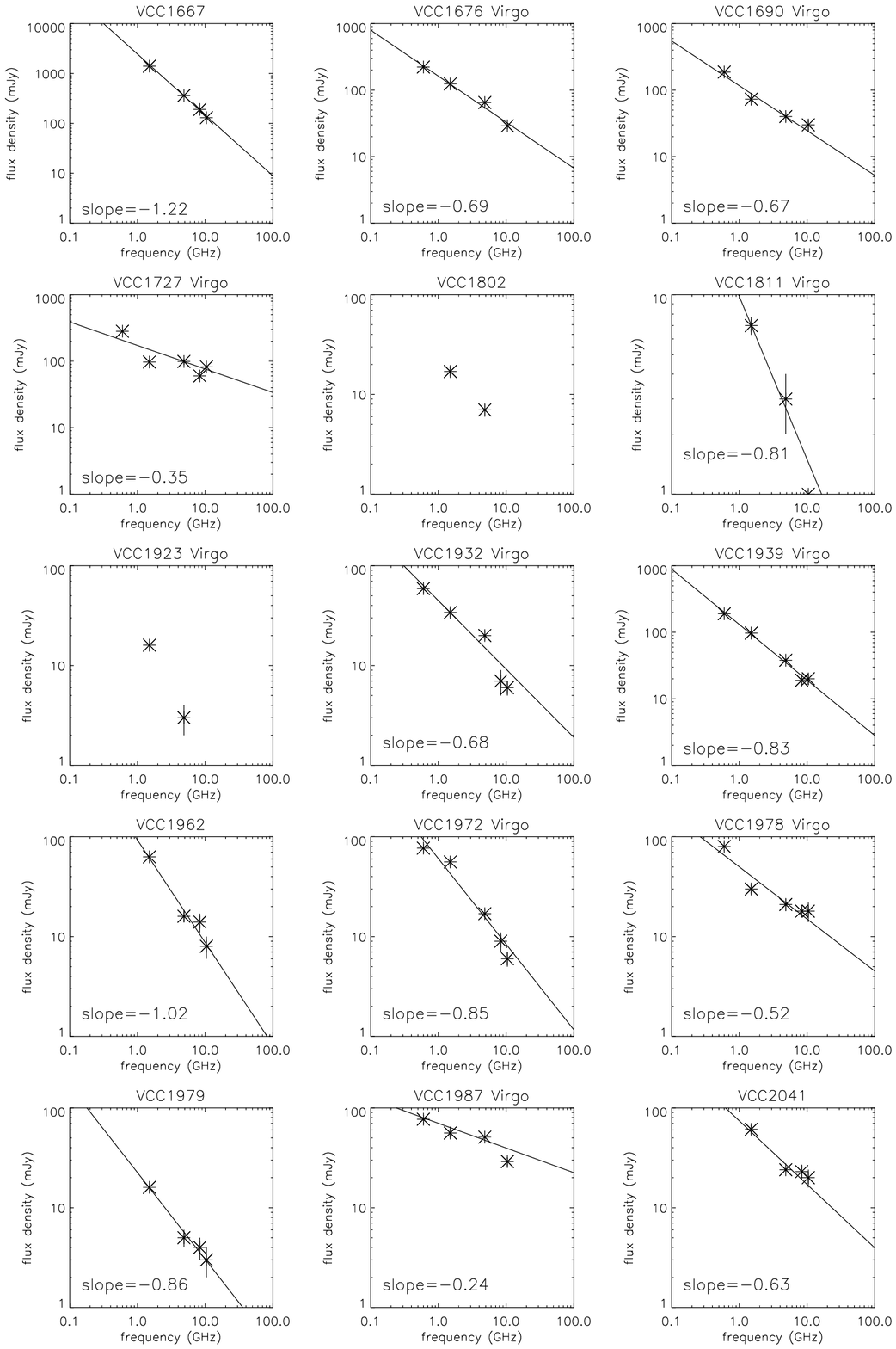}}
        \caption{Radio spectra of the observed VCC sources (continued). }
\end{figure*}
\addtocounter{figure}{-1}
\begin{figure*}
        \resizebox{14cm}{!}{\includegraphics{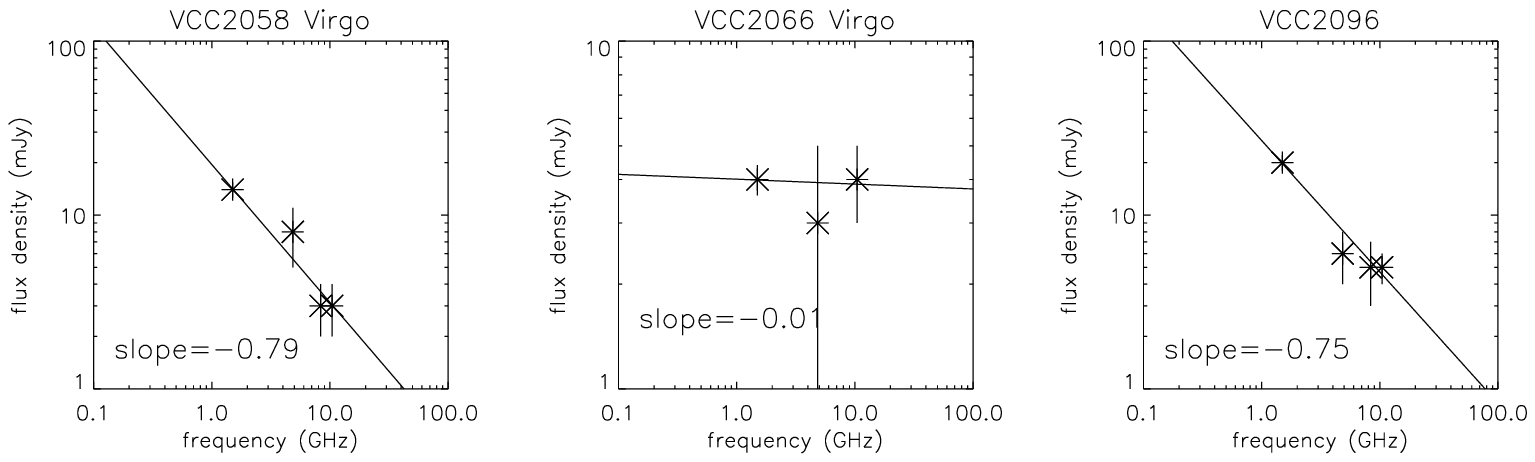}}
        \caption{Radio spectra of the observed VCC sources (continued). }
\end{figure*}

\section{The radio--FIR correlation \label{sec:radiofir}}

\begin{figure*}
        \resizebox{\hsize}{!}{\includegraphics{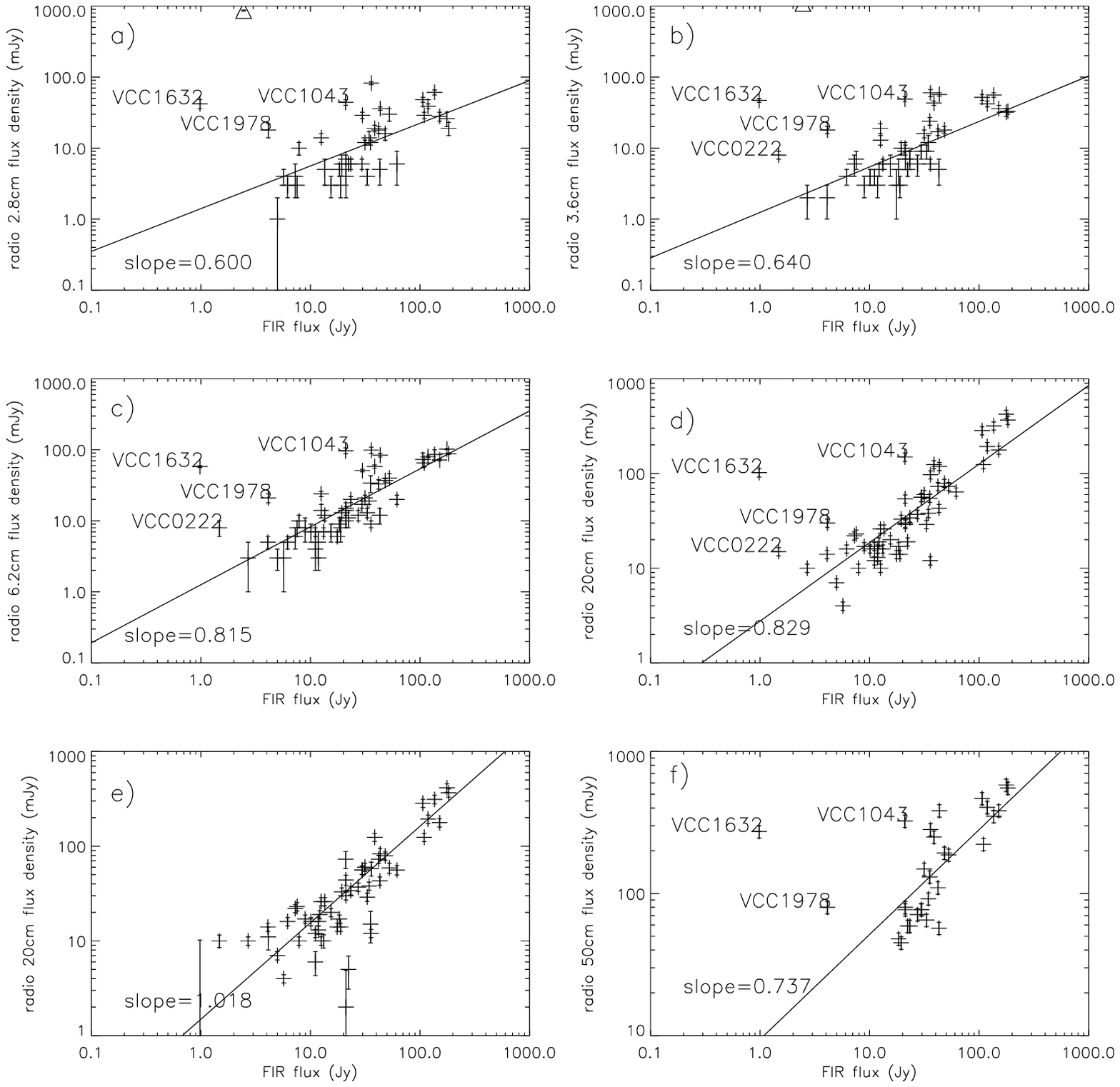}}
        \caption{Radio--FIR correlation: (a) at 10.55~GHz, (b) at 8.6~GHz, (c) at 4.85~GHz,
(d) at 1.4~GHz (NVSS) (e) at 1.4~GHz FIRST central point source flux densities subtracted,
(f) at 0.6~GHz.
        } \label{fig:radio_fir}
\end{figure*}

In Fig.~\ref{fig:radio_fir}(a)--(d) and (f) we show the radio--FIR correlation for
radio frequencies 10.55, 8.6, 4.85, 1.4, and 0.6~GHz including all galaxies of the sample.
The early type galaxies (E) of our sample are marked as triangles. Only one elliptical
appears in Fig.~\ref{fig:radio_fir}(a) and (b), because there are no FIR flux densities measured
for the other Virgo early type galaxies. Thus, the sample we use here is made of spiral (Sa-Im)
and S0 galaxies except VCC0763. The least absolute deviation
fits are also shown as solid lines whose slopes are plotted in the lower left corners
of the boxes.  The major outlying galaxies are labeled.
In order to investigate if the radio excess of these outliers is located in the disk
or in the center of the galaxies, we subtracted the FIRST 1.4~GHz flux densities
for sources where the FIRST image shows a central point source (the FIRST resolution is $6''$).
The radio--FIR correlation using these corrected 1.4~GHz flux densities can be seen
in Fig.~\ref{fig:radio_fir}(e). Clearly, the radio excesses are due to central point
sources, i.e central activity (LINER, HII, AGN). The slopes of the radio--FIR correlation at
4.85 and 1.4~GHz are consistent with those of samples of isolated galaxies (Wunderlich et al. 1987,
Niklas 1997). At 10.55~GHz our radio data is quite noisy, but when one removes the outlying points
the slope is consistent with that found by Niklas et al. (1995) and Niklas (1997).
The radio--FIR correlations at 8.6 and 0.6~GHz fit also in this picture.

In order to investigate the location of the Virgo galaxies that show a strong absolute radio excess
within the cluster, Fig.~\ref{fig:radioexcess} shows the absolute radio excess at 4.85~GHz
$S_{\rm 4.85~GHz}-S_{\rm FIR}$ as a function of the projected distance to the cluster center (M87).
The radio excess is defined as the difference between the observed and the from the
radio-FIR correlation expected radio flux.
All shown galaxies are classified as spiral or S0 galaxies. We use the measurements at 4.85~GHz,
because they are our most accurate measurements. The 4 galaxies that have the highest
absolute radio excesses are located at projected
distances smaller than 2 degrees ($\sim 0.6$~Mpc) and are labeled
in Fig.~\ref{fig:radioexcess}. All four galaxies are classified as centrally active
(VCC0836 (NGC~4388): Sy2, VCC1043 (NGC~4388): Sy/LINER, VCC1632 (NGC~4552): LINER/HII,
VCC1727 (NGC~4579): Sy2/LINER).
In addition, VCC0836, VCC1043, and VCC1727 galaxies show extended radio emission.
In the case of VCC1043 the radio emission, which is not due to a jet, is even extraplanar (Kotanyi et al. 1983).
But as can be seen in Fig.~\ref{fig:radio_fir}(e), when one subtracts the flux density of the
FIRST central point sources, there is no more radio excess observed.
We have also calculated the relative radio excess, i.e. the absolute radio excess divided by the
expected radio flux density calculated using the radio-FIR correlation. All four galaxies cited above
have a relative radio excess greater than 2. Additionally, this is also the case for
VCC0222 (NGC~4235, Sy1) and VCC1978. Thus, only galaxies that host an active center show a radio excess.

\section{The spectral index distribution \label{sec:specindex}}

The distribution of the spectral index of the Virgo galaxies as a function of the projected distance
to the cluster center (M87) can be seen in Fig.~\ref{fig:slopes}.
Elliptical galaxies are represented as triangles. The galaxies with a spectral index
greater than -0.5 are labeled. They are all spiral or S0 galaxies.
VCC0222 is classified as a Sy1, VCC1110 as a LINER, and VCC1727
as a Sy2. The spectral index of VCC1200 has to be taken with some caution, because the 8.6~GHz flux density
might be overestimated. VCC1450, VCC1535 and VCC1987 do not show any particular central activity.
It might be worth noting that VCC1987 (NGC~4654) has most probably undergone a tidal interaction and is
experiencing ram pressure now (Vollmer 2003). Moreover, it shows an off-center molecular bar
(Sofue et al. 2003). So one could speculate that it is just developing
a central activity. The galaxies with a spectral index $> -0.5$ are not concentrated towards
the cluster center. Thus there is no indication that cluster environment leads to
flatter spectral indices via central activity nor to steeper spectra as expected from better
confinement of relativistic electrons.
There might be a weak trend that the dispersion of the spectral index increases with decreasing
projected distance to the cluster center. The steepest spectral indices $\alpha < -1.0$ are found
in Sc/Scd galaxies and the elliptical galaxy VCC0763.

The number distribution of the spectral index is shown in Fig.~\ref{fig:histslopes}.
The dashed line traces the distribution of all sample galaxies, the solid line that of the
Virgo cluster members. Both distributions peak around a spectral index of $\alpha = -0.9$.
Both distributions show a tail to larger spectral indices than the peak value.
The steepest spectra $\alpha < -1.2$ are found outside the Virgo cluster.

\begin{figure}
        \resizebox{\hsize}{!}{\includegraphics{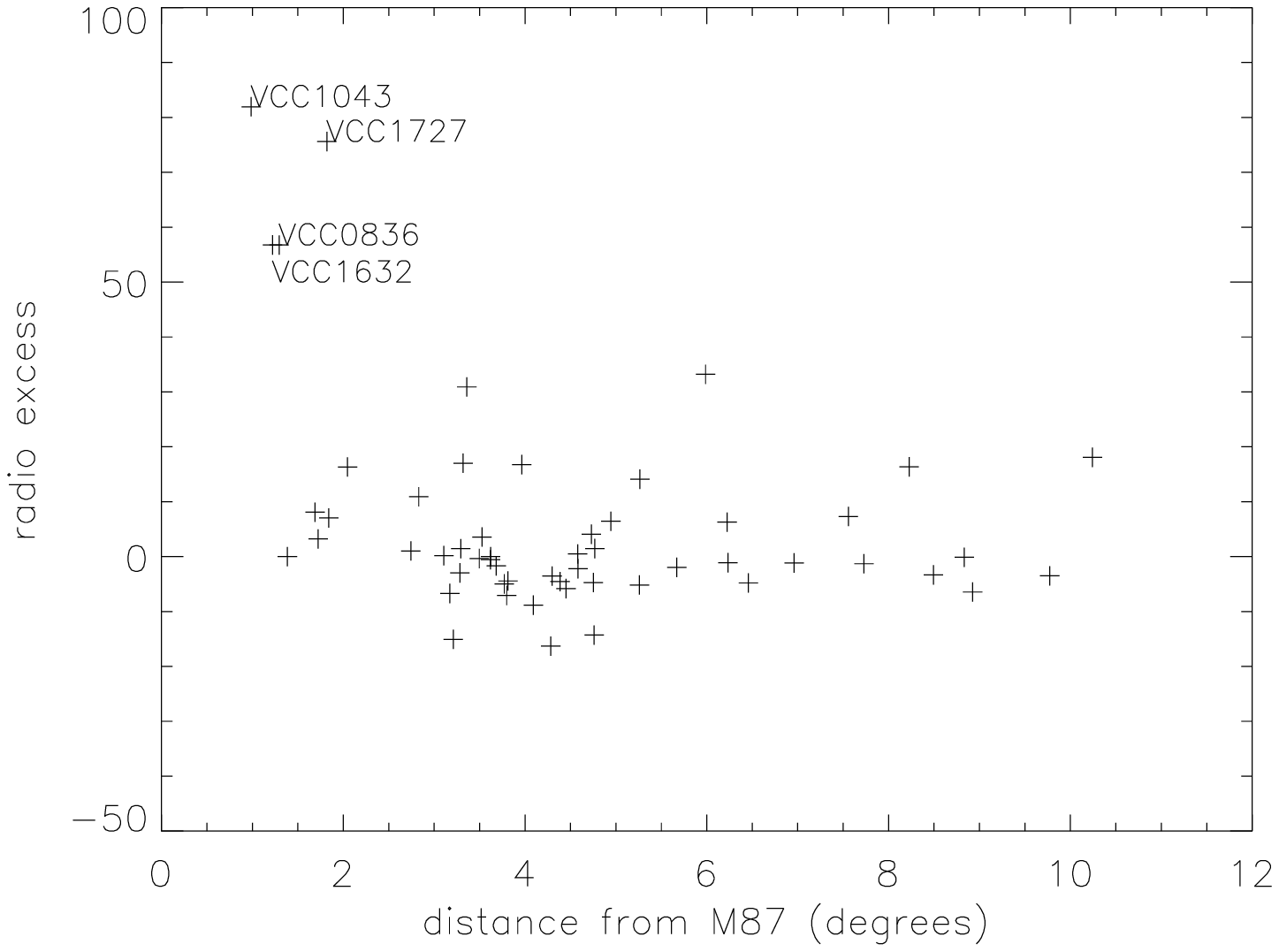}}
        \caption{Radio excess over the radio-FIR correlation of the observed VCC galaxies as a function
   of the projected distance to the cluster center (M87).
        } \label{fig:radioexcess}
\end{figure}

\begin{figure}
        \resizebox{\hsize}{!}{\includegraphics{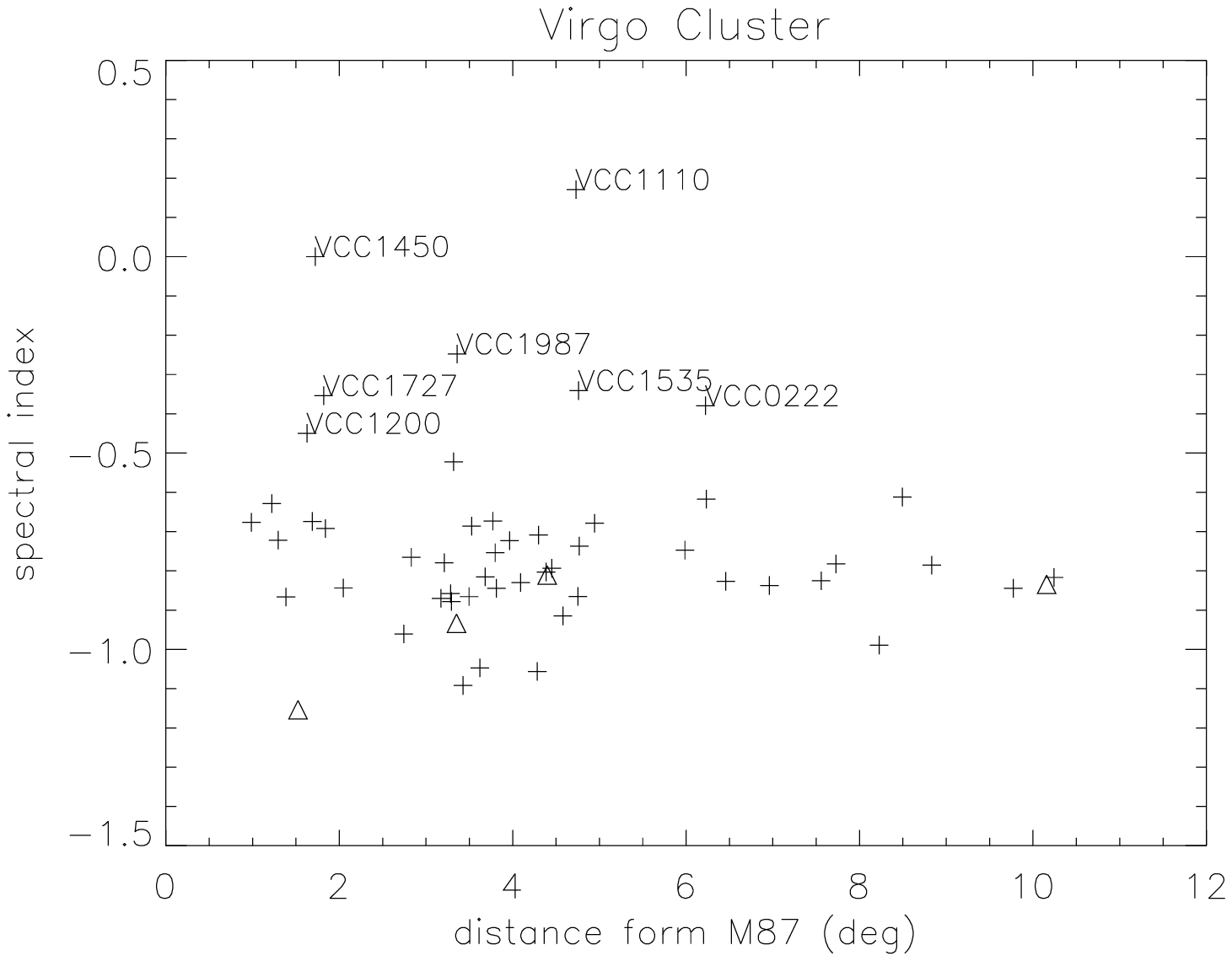}}
        \caption{Spectral index of the Virgo galaxies as a function of the projected distance to the
   cluster center (M87).
        } \label{fig:slopes}
\end{figure}

\begin{figure}
        \resizebox{\hsize}{!}{\includegraphics{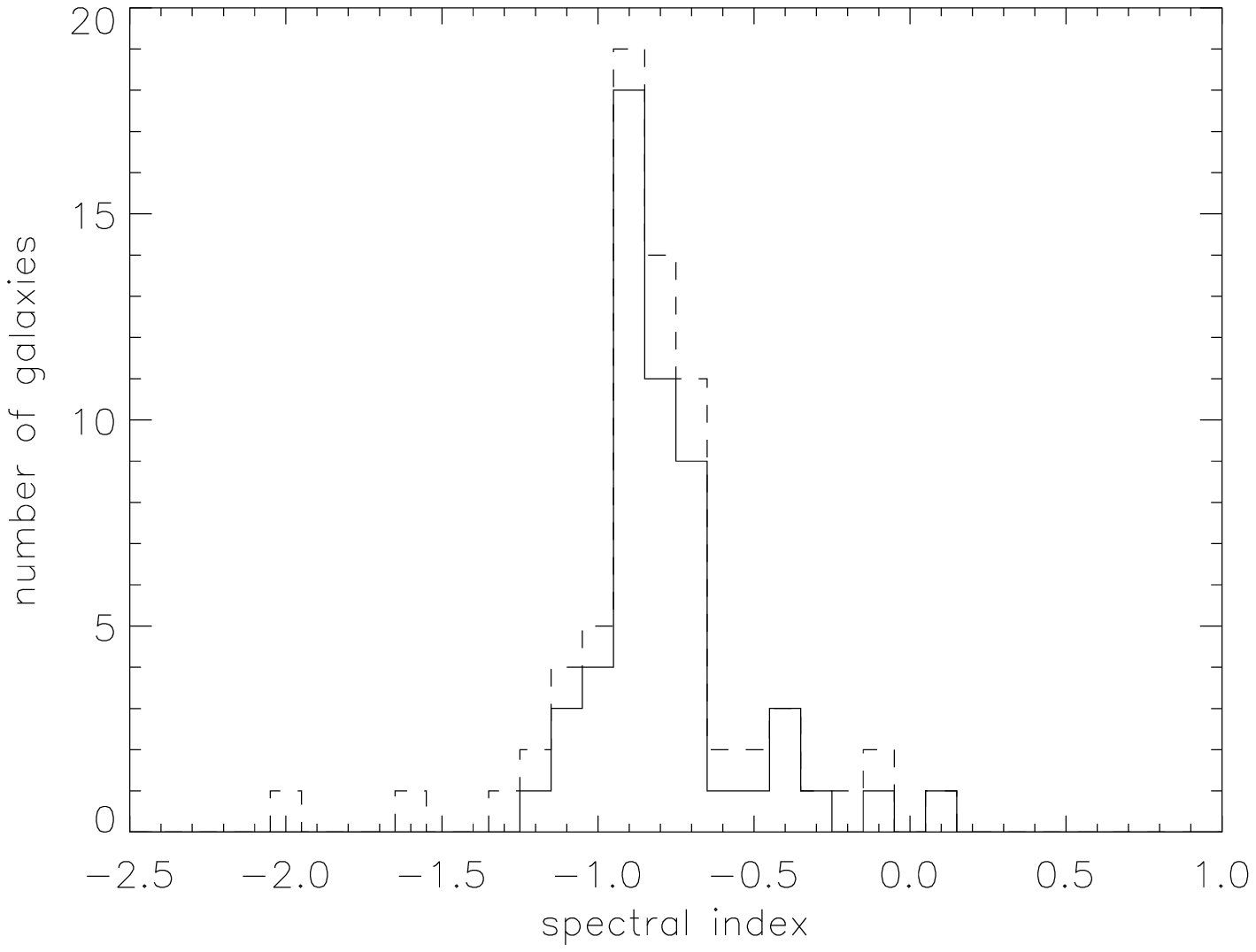}}
        \caption{Number distribution of the spectral index. Solid line: Virgo galaxies. Dashed line:
all sample galaxies.
        } \label{fig:histslopes}
\end{figure}

\section{Discussion and Conclusions \label{sec:discon}}

We confirm the finding of Gavazzi \& Boselli (1999b) that Virgo cluster galaxies statistically do not show
an enhanced radio emission compared to isolated galaxies.
However, a few spiral galaxies show an excess in
the absolute radio/FIR ratio. These galaxies all possess active centers (LINER, HII, AGN)
and are located within a radius of $2^{\rm o}$ from the cluster center.
We confirm the conclusion of Niklas et al. (1995) based on a small number of galaxies that
the spectral index of Virgo galaxies does not change significantly with projected cluster distance.
There might be a trend that the scatter of the spectral indices with decreasing projected distance.
The change in the behaviour of radio sources within the Virgo cluster (radio excess and
spectral index) is due to central activity. It is not clear if and under which conditions
the cluster environment triggers central activity.

As described in Sect.~\ref{sec:intro}, during the phase of active ram pressure stripping, the
magnetic field is compressed and one expects an excess of radio emission with respect to
the FIR emission. This extended radio emission should be asymmetric in the outer parts of the disk.
This seems not to be the case in the Virgo cluster. Based on the H{\sc i} observations
of the brightest spiral galaxies (Cayatte et al. 1990), no clear case for active ram pressure
stripping could be identified. The only possible case is VCC1043 (NGC~4438), but there
the galaxy had also a strong tidal interaction (Combes et al. 1988).
A clearer case is the edge-on galaxy VCC1516 (NGC~4522) where high column density extraplanar H{\sc i}
is detected (Kenney et al. 2003, in prep.). Both galaxies do not show an enhanced, asymmetric
radio continuum emission. However, in VCC1516 the compression is visible as a region of enhanced
polarization.
In barred galaxies the situation is similar. Whereas in the compression upstream regions
of the bar the radio polarization is enhanced, an excess of total radio continuum
emission is not observed (Beck et al. 2001).
We might speculate that this is the case, because both galaxies
are stripped almost face-on, in which case the asymmetric compression is minimum. Other possibilities
are: (i) the turbulent magnetic field is enhanced via an enhanced stars formation rate due to compression.
The enhancement of the ordered magnetic field due to compression might then be negligible
with respect to the turbulent magnetic field, but it can still be observed in polarization.
(ii) During compression the scale height of the magnetic field increases, which leads to
a decrease of the total magnetic field strength.

In the Coma cluster the situation is different.  H{\sc i} observations of the brightest spiral galaxies
revealed highly asymmetric gas disks of high column density (Bravo-Alfaro et al. 2000).
At the same time, Gavazzi \& Boselli (1999b) found a statistical excess of radio emissivity
for Coma cluster galaxies. This is a puzzling result, because the Virgo cluster is
dynamical young (asymmetric galaxy distribution, constant velocity distribution of spirals),
whereas the Coma cluster is more relaxed. One would thus expect that ram pressure is more
active in the Virgo cluster. The solution to this paradigm is that the core radius of the
Virgo cluster ICM distribution is much smaller than that of the Coma cluster.
Thus Virgo galaxies have to come very close to the cluster center to be significantly stripped
(Vollmer et al. 2001). In this case the phase of active ram pressure stripping is
much shorter in the Virgo cluster than in the Coma cluster due to the small ICM core radius and
the high galaxy velocities near the cluster center.

\renewcommand\baselinestretch{1}
\begin{table*}
      \caption{Radio and FIR fluxes of the galaxy sample.}
         \label{tab:parameters}
      \[
         \begin{array}{lllllllllllll}
         \hline
  \hline
             {\rm Name} & {\rm RA (B1950)} & {\rm DEC (B1950)} & {\rm D} & {\rm type} & {\rm Virgo} & {\rm S}$$_{2.8}^{\rm a}$$ & {\rm S}$$_{3.6}^{\rm a}$$ & {\rm S}$$_{6.2}^{\rm a}$$ & {\rm S}$$_{20}^{\rm b}$$ & {\rm S}$$_{50}^{\rm b}$$ & {\rm S}$$_{\rm FIR}^{\rm c}$$ & \alpha \\
                        &                  &                   & {\rm (deg.)} &       &             & {\rm (mJy)}               & {\rm (mJy)}               & {\rm (mJy)}               & {\rm (mJy)}              & {\rm (mJy)}              & {\rm (Jy)}                    &        \\
  \hline
  \hline
{\rm VCC}0025 & 12\ 08\ 04 & 16\ 18\ 39 & 6.23 & {\rm Sc } & {\rm y} & 6 \pm 2 & 10 \pm 2 & 13 \pm 2 & 33 & 45 & 19.35 & -0.61\\
{\rm VCC}0066 & 12\ 10\ 13 & 11\ 08\ 47 & 4.76 & {\rm Sc } & {\rm y} & 5 \pm 2 &  -  & 12 \pm 2 & 26 &  -  & 13.52 & -0.73 \\
{\rm VCC}0073 & 12\ 10\ 29 & 07\ 19\ 00 & 6.96 & {\rm Sb } & {\rm y} & 4 \pm 2 & 9 \pm 2 & 14 \pm 2 & 30 & 77 & 21.11 & -0.83 \\
{\rm VCC}0089 & 12\ 11\ 15 & 13\ 42\ 11 & 4.38 & {\rm Sc } & {\rm y} & 5 \pm 1 & 5 \pm 2 & 9 \pm 2 & 17 & 48 & 18.40 & -0.80 \\
{\rm VCC}0092 & 12\ 11\ 15 & 15\ 10\ 32 & 4.94 & {\rm Sb } & {\rm y} & {\bf 18 \pm 2} & 17 \pm 2 & {\bf 33 \pm 5} & 73 & 110 & 41.99 & -0.67 \\
{\rm VCC}0120 & 12\ 12\ 05 & 06\ 05\ 06 & 7.72 & {\rm Scd } & {\rm y} &  -  & 4 \pm 1 & 7 \pm 2 & 16 &  -  & 10.09 & -0.78 \\
{\rm VCC}0152 & 12\ 12\ 57 & 09\ 51\ 45 & 4.75 & {\rm Scd } & {\rm y} & 3 \pm 1 & 6 \pm 2 & 7 \pm 2 & 20 &  -  & 15.41 & -0.86 \\
{\rm VCC}0154 & 12\ 13\ 01 & 14\ 13\ 14 & 4.12 & {\rm E/S0 } & {\rm n} & 12 \pm 2 & 16 \pm 3 & 19 \pm 3 & 34 &  -  &  - & -0.50 \\
{\rm VCC}0157 & 12\ 13\ 05 & 14\ 10\ 41 & 4.09 & {\rm Sc } & {\rm y} & 4 \pm 1 & 9 \pm 2 & 13 \pm 2 & 29 & 65 & 33.03 & -0.82 \\
{\rm VCC}0167 & 12\ 13\ 21 & 13\ 25\ 38 & 3.81 & {\rm Sb } & {\rm y} &  -  & 3 \pm 1 & 5 \pm 2 & 13 &  -  & 11.81 & -0.84 \\
{\rm VCC}0222 & 12\ 14\ 36 & 07\ 28\ 09 & 6.22 & {\rm Sa } & {\rm y} &  -  & 8 \pm 1 & 8 \pm 2 & 15 &  -  & 1.47 & -0.37 \\
{\rm VCC}0224 & 12\ 14\ 35 & 12\ 43\ 57 & 3.42 & {\rm Scd } & {\rm y} & 16 \pm 2 & 22 \pm 2 & 42 \pm 2 & 139 &  -  &  -  & -1.09 \\
{\rm VCC}0302 & 12\ 16\ 09 & 12\ 00\ 24 & 3.10 & {\rm Sc } & {\rm n} &  -  & 2 \pm 1 & 3 \pm 2 & 10 &  -  & 2.69 & -0.95 \\
{\rm VCC}0307 & 12\ 16\ 17 & 14\ 41\ 29 & 3.62 & {\rm Sc } & {\rm y} & 19 \pm 4 & 33 \pm 3 & 88 \pm 2 & 366 & 553 & 182.45 & -1.04 \\
{\rm VCC}0311 & 12\ 16\ 21 & 13\ 11\ 25 & 3.03 & {\rm E/S0 } & {\rm n} &  -  & 9 \pm 3 & 12 \pm 2 & 12 &  -  &  -  & -0.07 \\
{\rm VCC}0382 & 12\ 17\ 22 & 05\ 37\ 16 & 7.55 & {\rm Sc } & {\rm y} & 16 \pm 3 & 18 \pm 2 & 37 \pm 2 & 79 & 193 & 48.16 & -0.82 \\
{\rm VCC}0403 & 12\ 17\ 44 & 10\ 36\ 08 & 3.35 & {\rm dE } & {\rm y} & 13 \pm 4 & 19 \pm 2 & 33 \pm 2 & 94 &  -  &  -  & -0.93 \\
{\rm VCC}0414 & 12\ 17\ 51 & 14\ 58\ 12 & 3.48 & {\rm dE } & {\rm y} &  -  & 6 \pm 1 &  -  & 14 &  -  &  - &  \\
{\rm VCC}0435 & 12\ 18\ 15 & 17\ 17\ 36 & 5.26 & {\rm S } & {\rm n} &  -  & 19 \pm 3 & 24 \pm 2 & 26 &  -  & 12.52 & -0.12 \\
{\rm VCC}0437 & 12\ 18\ 15 & 17\ 46\ 12 & 5.68 & {\rm dE } & {\rm y} &  -  &  -  & 5 \pm 1 & 15 &  -  &  - &  \\
{\rm VCC}0460 & 12\ 18\ 40 & 18\ 39\ 34 & 6.45 & {\rm Sa } & {\rm y} & 6 \pm 1 & 5 \pm 1 & 11 \pm 3 & 19 & 59 & 22.20 & -0.82 \\
{\rm VCC}0483 & 12\ 19\ 00 & 14\ 52\ 58 & 3.21 & {\rm Sc } & {\rm y} & 5 \pm 2 & 5 \pm 2 & 12 \pm 3 & 43 & 57 & 42.99 & -0.77 \\
{\rm VCC}0491 & 12\ 19\ 08 & 11\ 46\ 39 & 2.45 & {\rm Scd } & {\rm y} &  -  & 5 \pm 1 &  -  & 19 &  -  & 12.17 &  \\
{\rm VCC}0497 & 12\ 19\ 10 & 14\ 52\ 22 & 3.17 & {\rm Sc } & {\rm y} &  -  & 6 \pm 2 & 12 \pm 2 & 37 & 71 & 27.34 & -0.87 \\
{\rm VCC}0508 & 12\ 19\ 21 & 04\ 45\ 03 & 8.22 & {\rm Sc } & {\rm y} & 26 \pm 6 & 32 \pm 4 & 102 \pm 2 & 423 & 580 & 176.47 & -0.98 \\
{\rm VCC}0559 & 12\ 19\ 59 & 15\ 48\ 54 & 3.77 & {\rm Sab } & {\rm y} &  -  & 4 \pm 1 & 4 \pm 2 & 12 &  -  & 11.10 & -0.67 \\
{\rm VCC}0572 & 12\ 20\ 09 & 03\ 25\ 06 & 9.47 & {\rm E/S0 } & {\rm n} &  -  & 7 \pm 1 & 6 \pm 2 & 14 &  -  &  - & -0.42 \\
{\rm VCC}0580 & 12\ 20\ 12 & 12\ 34\ 15 & 2.02 & {\rm BCD } & {\rm n} &  -  & 4 \pm 2 & 5 \pm 1 & 21 &  -  &  -  & -1.13 \\
{\rm VCC}0596 & 12\ 20\ 22 & 16\ 06\ 01 & 3.96 & {\rm Sc } & {\rm y} & {\bf 61 \pm 5} & 56 \pm 4 & {\bf 86 \pm 14} & 318 & 350 & 136.00 & -0.72 \\
{\rm VCC}0638 & 12\ 20\ 50 & 07\ 45\ 00 & 5.25 & {\rm Sab } & {\rm n} & 7 \pm 1 & 7 \pm 2 & 10 \pm 2 & 29 &  -  & 21.11 & -0.76 \\
{\rm VCC}0763 & 12\ 22\ 31 & 13\ 09\ 12 & 1.52 & {\rm E } & {\rm y} & 864 \pm 15 & 1085 \pm 20 & 2270 \pm 20 & 6067 & 10802 & 2.45 & -1.15 \\
{\rm VCC}0801 & 12\ 22\ 53 & 16\ 44\ 47 & 4.29 & {\rm Sa } & {\rm y} & 14 \pm 3 & 12 \pm 3 & 19 \pm 2 & 38 & 92 & 34.36 & -0.70 \\
{\rm VCC}0831 & 12\ 23\ 10 & 09\ 18\ 29 & 3.59 & {\rm Sc } & {\rm n} & 5 \pm 1 & 8 \pm 2 & 12 \pm 2 & 25 &  -  &  - & -0.74 \\
{\rm VCC}0836 & 12\ 23\ 14 & 12\ 56\ 20 & 1.29 & {\rm Sab } & {\rm y} & {\bf 36 \pm 2} & 57 \pm 3 & {\bf 84 \pm 5} & 119 & 383 & 43.32 & -0.72 \\
{\rm VCC}0865 & 12\ 23\ 27 & 15\ 56\ 49 & 3.49 & {\rm Sc } & {\rm y} & 4 \pm 2 & 6 \pm 2 & 6 \pm 2 & 22 &  -  & 7.26 & -0.86 \\
{\rm VCC}0873 & 12\ 23\ 34 & 13\ 23\ 24 & 1.38 & {\rm Sc } & {\rm y} & {\bf 12 \pm 1} & 16 \pm 2 & {\bf 21 \pm 2} & 60 & 149 & 31.48 & -0.86 \\
{\rm VCC}0904 & 12\ 23\ 55 & 09\ 17\ 54 & 3.54 & {\rm Pec } & {\rm n} & 9 \pm 3 & 12 \pm 4 & 42 \pm 2 & 404 &  -  &  - & -1.94 \\
{\rm VCC}0907 & 12\ 23\ 55 & 09\ 17\ 49 & 3.54 & {\rm Pec } & {\rm n} & 11 \pm 3 & 11 \pm 3 & 45 \pm 2 & 257 &  -  &  - & -1.56 \\
{\rm VCC}0921 & 12\ 24\ 02 & 04\ 14\ 29 & 8.49 & {\rm Sbc } & {\rm y} &  -  & 6 \pm 1 & 7 \pm 1 & 16 &  -  & 13.18 & -0.61 \\
{\rm VCC}0958 & 12\ 24\ 24 & 15\ 19\ 28 & 2.82 & {\rm Sa } & {\rm y} & {\bf 12 \pm 1} & 24 \pm 3 & {\bf 34 \pm 9} & 55 & 131 & 35.38 & -0.76 \\
{\rm VCC}1040 & 12\ 25\ 10 & 13\ 15\ 22 & 0.97 & {\rm dE } & {\rm y} &  -  & 3 \pm 1 &  -  & 17 &  -  &  -  &  \\
{\rm VCC}1043 & 12\ 25\ 13 & 13\ 17\ 07 & 0.98 & {\rm Sb } & {\rm y} & {\bf 44 \pm 4} & 49 \pm 4 & {\bf 97 \pm 9} & 149 & 324 & 20.97 & -0.67 \\
{\rm VCC}1110 & 12\ 25\ 58 & 17\ 21\ 38 & 4.72 & {\rm Sab } & {\rm y} & {\bf 14 \pm 2} & 13 \pm 2 & 14 \pm 3 & 10 &  -  & 12.55 & \phminus 0.17 \\
{\rm VCC}1145 & 12\ 26\ 25 & 03\ 50\ 49 & 8.83 & {\rm Sb } & {\rm y} &  -  & 10 \pm 2 & 15 \pm 3 & 54 & 80 & 21.06 & -0.78 \\
{\rm VCC}1200 & 12\ 27\ 02 & 11\ 04\ 10 & 1.62 & {\rm Im } & {\rm y} &  -  & 4 \pm 2 & 9 \pm 2 & 13 &  -  &  -  & -0.44 \\
{\rm VCC}1205 & 12\ 27\ 05 & 08\ 05\ 53 & 4.57 & {\rm Sc } & {\rm y} &  -  & 3 \pm 1 & 8 \pm 3 & 17 &  -  & 8.94 & -0.91 \\
{\rm VCC}1226 & 12\ 27\ 14 & 08\ 16\ 39 & 4.39 & {\rm E/S0 } & {\rm y} & 42 \pm 3 & 52 \pm 3 & 84 \pm 3 & 248 & 392 &  - & -0.81 \\
{\rm VCC}1376 & 12\ 29\ 07 & 04\ 12\ 13 & 8.46 & {\rm Sc } & {\rm n} & 6 \pm 2 & 4 \pm 1 & 10 \pm 3 & 17 &  -  &  - & -0.68 \\
{\rm VCC}1393 & 12\ 29\ 24 & 15\ 23\ 46 & 2.74 & {\rm Sc } & {\rm y} &  -  & 2 \pm 1 & 5 \pm 1 & 14 &  -  & 4.08 & -0.96 \\
{\rm VCC}1401 & 12\ 29\ 27 & 14\ 41\ 38 & 2.04 & {\rm Sbc } & {\rm y} & {\bf 48 \pm 4} & 52 \pm 5 & {\bf 73 \pm 17} & 284 & 467 & 106.37 & -0.84 \\
{\rm VCC}1413 & 12\ 29\ 35 & 12\ 41\ 59 & 0.32 & {\rm dE } & {\rm y} &  -  &  -  &  -  & 43 &  -  &  - &  \\
{\rm VCC}1449 & 12\ 30\ 08 & 13\ 35\ 35 & 1.03 & {\rm dE } & {\rm y} &  -  &  -  &  -  & 20 &  -  &  - &  \\
 \hline
        \end{array}
      \]
\begin{list}{}{}
\item[$^{\rm{a}}$] this paper and measurements of Niklas et al. (1995) in boldface
\item[$^{\rm{b}}$] from Goldmine (Gavazzi et al. 2003)
\end{list}
\end{table*}

\addtocounter{table}{-1}
\begin{table*}
      \caption{Radio and FIR fluxes of the galaxy sample (continued).}
      \[
         \begin{array}{lllllllllllll}
         \hline
  \hline
             {\rm Name} & {\rm RA (B1950)} & {\rm DEC (B1950)} & {\rm D} & {\rm type} & {\rm Virgo} & {\rm S}$$_{2.8}^{\rm a}$$ & {\rm S}$$_{3.6}^{\rm a}$$ & {\rm S}$$_{6.2}^{\rm a}$$ & {\rm S}$$_{20}^{\rm b}$$ & {\rm S}$$_{50}^{\rm b}$$ & {\rm S}$$_{\rm FIR}^{\rm c}$$ & \alpha \\
                        &                  &                   & {\rm (deg.)} &       &             & {\rm (mJy)}               & {\rm (mJy)}               & {\rm (mJy)}               & {\rm (mJy)}              & {\rm (mJy)}              & {\rm (Jy)}                    &        \\
  \hline
  \hline
{\rm VCC}1450 & 12\ 30\ 10 & 14\ 19\ 35 & 1.72 & {\rm Sc } & {\rm y} & {\bf 10 \pm 2} &  -  & 10 \pm 2 & 10 &  -  & 7.87 & \phminus 0.00 \\
{\rm VCC}1508 & 12\ 30\ 58 & 08\ 55\ 43 & 3.79 & {\rm Sc } & {\rm y} &  -  & 3 \pm 2 & 6 \pm 1 & 14 &  -  & 17.60 & -0.75 \\
{\rm VCC}1516 & 12\ 31\ 07 & 09\ 27\ 03 & 3.29 & {\rm Sbc } & {\rm y} & 3 \pm 1 & 7 \pm 2 & 8 \pm 2 & 23 &  -  & 7.55 & -0.87 \\
{\rm VCC}1535 & 12\ 31\ 30 & 07\ 58\ 30 & 4.76 & {\rm S0 } & {\rm y} &  -  & 6 \pm 1 & 9 \pm 1 & 12 &  -  & 35.75 & -0.34 \\
{\rm VCC}1540 & 12\ 31\ 35 & 02\ 55\ 46 & 9.77 & {\rm Sb } & {\rm y} & 28 \pm 4 & 36 \pm 4 & 72 \pm 2 & 177 & 383 & 151.21 & -0.84 \\
{\rm VCC}1554 & 12\ 31\ 46 & 06\ 44\ 40 & 5.98 & {\rm Sm } & {\rm y} & {\bf 19 \pm 2} & 43 \pm 3 & {\bf 58 \pm 3} & 124 & 250 & 38.60 & -0.74 \\
{\rm VCC}1555 & 12\ 31\ 47 & 08\ 28\ 25 & 4.28 & {\rm Sc } & {\rm y} & 6 \pm 3 &  -  & 20 \pm 3 & 64 &  -  & 61.62 & -1.05 \\
{\rm VCC}1562 & 12\ 31\ 53 & 02\ 27\ 50 & 10.2 & {\rm Sc } & {\rm y} & 39 \pm 3 & 42 \pm 4 & 80 \pm 2 & 193 & 404 & 118.57 & -0.81 \\
{\rm VCC}1632 & 12\ 33\ 08 & 12\ 49\ 53 & 1.22 & {\rm S0 } & {\rm y} & 42 \pm 6 & 47 \pm 3 & 58 \pm 2 & 102 & 273 & 0.98 & -0.62 \\
{\rm VCC}1658 & 12\ 33\ 45 & 13\ 51\ 07 & 1.80 & {\rm dE } & {\rm y} &  -  &  -  & 3 \pm 2 & 11 &  -  &  - &  \\
{\rm VCC}1667 & 12\ 33\ 58 & 16\ 49\ 02 & 4.38 & {\rm dE } & {\rm n} & 130 \pm 8 & 190 \pm 10 & 360 \pm 10 & 1402 &  -  &  - & -1.22 \\
{\rm VCC}1673 & 12\ 33\ 59 & 11\ 32\ 01 & 1.81 & {\rm Sc } & {\rm y} &  -  &  -  &  -  & 10 &  -  &  - & - \\
{\rm VCC}1676 & 12\ 34\ 02 & 11\ 30\ 55 & 1.84 & {\rm Sc } & {\rm y} & {\bf 29 \pm 3} &  -  & {\bf 65 \pm 8} & 124 & 222 & 109.33 & -0.69 \\
{\rm VCC}1690 & 12\ 34\ 18 & 13\ 26\ 27 & 1.68 & {\rm Sab } & {\rm y} & {\bf 30 \pm 6} &  -  & {\bf 40 \pm 6} & 73 & 187 & 52.60 & -0.67 \\
{\rm VCC}1727 & 12\ 35\ 12 & 12\ 05\ 37 & 1.81 & {\rm Sab } & {\rm y} & {\bf 82 \pm 4} & 60 \pm 7 & {\bf 99 \pm 10} & 97 & 282 & 35.95 & -0.35 \\
{\rm VCC}1802 & 12\ 37\ 06 & 07\ 26\ 36 & 5.66 & {\rm S } & {\rm n} &  -  &  -  & 7 \pm 1 & 17 &  -  & 11.09 &  \\
{\rm VCC}1811 & 12\ 37\ 21 & 15\ 34\ 20 & 3.68 & {\rm Sc } & {\rm y} & {\bf 1 \pm 1} &  -  & 3 \pm 1 & 7 &  -  & 4.99 & -0.81 \\
{\rm VCC}1923 & 12\ 39\ 59 & 04\ 14\ 05 & 8.92 & {\rm Sbc } & {\rm y} &  -  &  -  & 3 \pm 1 & 16 &  -  & 11.83 &  \\
{\rm VCC}1932 & 12\ 40\ 10 & 14\ 34\ 12 & 3.52 & {\rm Sc } & {\rm y} & {\bf 6 \pm 1} & 7 \pm 2 & {\bf 20 \pm 2} & 34 & 59 & 23.30 & -0.68 \\
{\rm VCC}1939 & 12\ 40\ 17 & 02\ 57\ 43 & 10.1 & {\rm E/S0 } & {\rm y} & 20 \pm 4 & 19 \pm 2 & 38 \pm 3 & 97 & 190 &  - & -0.83 \\
{\rm VCC}1962 & 12\ 40\ 45 & 03\ 49\ 26 & 9.37 & {\rm S0 } & {\rm n} & 8 \pm 2 & 14 \pm 3 & 16 \pm 2 & 63 &  -  &  - & -1.02 \\
{\rm VCC}1972 & 12\ 41\ 01 & 11\ 51\ 23 & 3.28 & {\rm Sc } & {\rm y} & {\bf 6 \pm 1} & 9 \pm 2 & 17 \pm 2 & 56 & 77 & 29.58 & -0.85 \\
{\rm VCC}1978 & 12\ 41\ 08 & 11\ 49\ 34 & 3.31 & {\rm S0 } & {\rm y} & 18 \pm 4 & 18 \pm 2 & 21 \pm 3 & 30 & 80 & 4.11 & -0.52 \\
{\rm VCC}1979 & 12\ 41\ 14 & 11\ 02\ 37 & 3.61 & {\rm Sc } & {\rm n} & 3 \pm 1 & 4 \pm 1 & 5 \pm 1 & 16 &  -  & 6.18 & -0.86 \\
{\rm VCC}1987 & 12\ 41\ 24 & 13\ 24\ 10 & 3.35 & {\rm Sc } & {\rm y} & {\bf 29 \pm 3} &  -  & {\bf 51 \pm 2} & 56 & 77 & 29.84 & -0.24 \\
{\rm VCC}2041 & 12\ 44\ 02 & 12\ 10\ 09 & 3.96 & {\rm E/S0 } & {\rm n} & 20 \pm 4 & 23 \pm 4 & 24 \pm 2 & 61 &  -  &  - & -0.63 \\
{\rm VCC}2058 & 12\ 45\ 14 & 14\ 02\ 05 & 4.45 & {\rm Sc } & {\rm y} & {\bf 3 \pm 1} & 3 \pm 1 & {\bf 8 \pm 3} & 14 &  -  & 18.87 & -0.79 \\
{\rm VCC}2066 & 12\ 45\ 44 & 11\ 15\ 22 & 4.58 & {\rm S0 } & {\rm y} & {\bf 4 \pm 1} &  -  & {\bf 3 \pm 2} & 4 &  -  & 5.69 & -0.01 \\
{\rm VCC}2096 & 12\ 50\ 54 & 11\ 58\ 54 & 5.69 & {\rm BCD } & {\rm n} & {\bf 5 \pm 1} & 5 \pm 2 & {\bf 6 \pm 2} & 20 &  -  &  - & -0.75 \\
        \hline
        \end{array}
      \]
\begin{list}{}{}
\item[$^{\rm{a}}$] this paper and measurements of Niklas et al. (1995) in boldface
\item[$^{\rm{b}}$] from Goldmine (Gavazzi et al. 2003)
\end{list}
\end{table*}
\renewcommand\baselinestretch{1.5}

\end{document}